\documentclass[preprint,12pt]{aastex}
\usepackage{epsfig}
\usepackage{graphicx}
\newcommand{\chandra}{{\it Chandra}}

\newcommand{\flux}{\thinspace\hbox{$\hbox{ergs}\thinspace\hbox{cm}^{-2}\thinspace\hbox{s}^{-1}$}}
\newcommand{\hst}{{\it HST}}

\makeatletter

\begin{document}

\def\spose#1{\hbox to 0pt{#1\hss}}
\def\laeq{\mathrel{\spose{\lower 3pt\hbox{$\mathchar"218$}}
     \raise 2.0pt\hbox{$\mathchar"13C$}}}
\def\gaeq{\mathrel{\spose{\lower 3pt\hbox{$\mathchar"218$}}
     \raise 2.0pt\hbox{$\mathchar"13E$}}}

\slugcomment{ }

\title{X-ray and Optical Study of Low Core Density
Globular Clusters NGC6144 and E3} 

\author{Shih Hao Lan\altaffilmark{1}, Albert~K.~H.~Kong\altaffilmark{1,8}, Frank~Verbunt\altaffilmark{2}, Walter~H.~G.~Lewin\altaffilmark{3},
Cees Bassa\altaffilmark{4,5}, Scott~F.~Anderson\altaffilmark{6}, 
 and David~Pooley\altaffilmark{7} 
}
\altaffiltext{1}{Institute of Astronomy and Department of Physics,
 National Tsing Hua University
  101 Section 2 Kuang Fu Road
  , Hsinchu, 
  Taiwan 30013,
  R.O.C.}

\altaffiltext{2}{Astronomical Institute, Utrecht University, P.O. Box 
80000, 3508 TA, Utrecht, the Netherlands}
\altaffiltext{3}{Kavli Institute for Astrophysics and Space Research,
Massachusetts Institute of Technology, 77
Massachusetts Avenue, Cambridge, MA 02139}
\altaffiltext{4}{SRON, Netherlands Institute for Space Research Sorbonnelaan 2, 3584 CA, Utrecht, The Netherlands}
\altaffiltext{5}{Department of Astrophysics, IMAPP, Radboud University Nijmegen PO Box 9010, 6500 GL Nijmegen, The Netherlands}
\altaffiltext{6}{Department of Astronomy, University of Washington, Box 
351580, Seattle, WA 98195}

\altaffiltext{7}{Astronomy Department, University of Wisconsin-Madison,  
475 North Charter Street, Madison, WI 53706 }

\altaffiltext{8}{Kenda Foundation Golden Jade Fellow}

\begin{abstract}

We report on the {\it Chandra X-ray Observatory} and {\it Hubble Space Telescope} observation of two low core density globular clusters, NGC6144 and E3. By comparing the number of X-ray sources inside the half-mass radius to those outside, we found 6 X-ray sources within the half-mass radius of NGC6144, among which 4 are expected to be background sources; 3 X-ray sources are also found within the half-mass radius of E3, of which 3 is expected to be background source. 
Therefore, we cannot exclude that all our sources are background sources.
However, combining the results from X-ray and optical observations, we found that 1-2 sources in NGC6144 and 1 source in E3 are likely to be cataclysmic variables and that 1 source in NGC6144 is an active binary, based on the X-ray and optical properties. The number of faint X-ray sources in NGC6144 and E3 found with \chandra\ and \hst\ is higher than a prediction based on collision frequency, but is closer to that based on mass. Our observations strongly suggest that the compact binary systems in NGC6144 and E3 are primordial in origin.

\end{abstract}


\section{Introduction}
The number density of X-ray sources in globular clusters with luminosities greater than $L_X = 10^{36}$ erg s$^{-1}$ is ~100 to ~1000 times higher than that of the Galactic disk (Giacconi et al. 1972;  Clark et al. 1975). With the improvement of X-ray instruments, sources with X-ray luminosities lower than $10^{34.5}$ erg s$^{-1}$ were discovered with the Einstein Observatory (Hertz and Grindlay 1983) in 1980s. The study of X-ray sources in Galactic globular clusters with the \chandra\ X-ray Observatory can reach luminosities down to $L_X=10^{29-30}$ erg s$^{-1}$ in the 0.5-6 keV band. The most luminous X-ray sources in globular clusters are neutron stars 
accreting matter from a companion star, and some dim X-ray sources with luminosities $L_X >$ $10^{32}$ erg s$^{-1}$ may be neutron stars in quiescence. According to previous research, dim X-ray sources with X-ray luminosities ranging from $10^{30}$ erg s$^{-1}$ to $10^{32}$ erg $s^{-1}$ are roughly identified as accreting white dwarfs or X-ray active low-mass main sequence star binaries (see e.g., Grindlay et al. 2001; Pooley et al. 2002, 2003; Bassa et al. 2004; Heinke et al. 2005; Verbunt \& Lewin 2006).

The high stellar densities in globular cluster cores make secure optical identifications of X-ray sources difficult. However, the high spatial resolution of \chandra\ allows us to determine X-ray source positions to sub-arcsecond levels, which can be used to identify optical counterparts in Hubble Space Telescope (HST) observations of the densest regions of globular clusters. Detailed identifications of low X-ray luminosity sources by using multi-wavelength approach has been applied in some nearby globular clusters and some globular clusters with low central density  (e.g. NGC 6397 by Grindlay et al. 2001, 47 Tuc by Heinke et al. 2005, M4 by Bassa et al. 2004, NGC 288 by Kong et al. 2006, NGC2808 by Servillat et al. 2008, M55 $\&$ NGC6366 by Bassa et al. 2008 and M12 by Lu et al. 2009). These studies suggest that the dim sources in globular clusters are quiescent low-mass X-ray binaries (qLMXBs; Wijnands et al. 2005 and references therein), cataclysmic variables (CVs; Cool et al. 1995; Bailyn et al. 1990), chromospherically active binaries (AB; Cool et al. 2002) or millisecond pulsars (MSPs; Grindlay et al. 2002). 

CVs consist of two components: a white dwarf and a main sequence or sub-giant star. Because of the accretion on the white dwarf, CVs are bluer than main sequence stars in the optical, and have higher H$\alpha$ and ultraviolet emission. Quiescent LMXBs are binaries with a neutron star but the accretion rates are much lower than in persistent LMXB. The optical colors of qLMXBs are between CVs and main sequence. Chromospherically ABs have strong coronal activity because of faster rotation periods. MSPs are neutron stars that emit radio emission with millisecond periods, and often have white dwarf or main sequence companions.

The study of Verbunt(2002) suggested that CVs are formed from close
encounters while ABs from primordial binaries. The study of Pooley et
al.(2003; see also Heinke et al. 2003) presents conclusive evidence of
the link between the number of close binaries observed in X-rays (
$L_{X(0.5-6keV)}>$ $4\times10^{30}$ ergs s$^{-1}$ ) and the encounter
rate of the globular cluster. It also suggests that CVs are mostly
formed from stellar encounters. Theoretical results of Davies (1997) also show 
that the progenitor binaries of CVs from wide primordial binaries are destroyed in the cores of globular clusters
with high core densities ($>10^3$ stars $pc^{-3}$).

Binary stars play important roles in star and globular cluster formation and evolution. 
The binding energy of these systems may be greater than that of the whole cluster. The dynamical process and the energy transfer of the binaries and the stars can strongly affect the dynamical evolution of the globular cluster (Hills 1975; Hills 1976; Pryor et al. 1991).  
Among clusters with a low core density and low encounter rate $\Gamma$,
the number of faint X-ray sources found in NGC 288 (Kong et al. 2006) is higher than the prediction based on stellar encounter frequency (Verbunt 2003; Pooley et al. 2003), which has been speculated as the evidence for primordial binaries. 
Pooley and Hut (2006) describe the number of X-ray sources (N) in globular clusters using a combination of the collision number ($\Gamma$) and mass (M): N=a$\Gamma$+bM, where M is the cluster mass. The best solution for a predicted number of X-ray sources in a globular is given by a=1.2 and b=1.1 (when $\Gamma$ and M are normalized on the values for M4; Verbunt et al. 2007; Bassa et al. 2008).  

We report here on the \chandra\ and \hst\ observations of the low core density globular clusters NGC6144 ($\rho_c = 10^{2.23} L_\odot \mathrm{pc}^{-3}$) and E3 ($\rho_c = 10^{1.11} L_\odot \mathrm{pc}^{-3}$; Harris, W.E. 1996); the \chandra\ observation and data analysis are described in \S2 , the \hst\ observation in \S3 , sources identification in \S4 and discussion in \S5 .

\section{X-RAY OBSERVATIONS AND DATA ANALYSIS}
The galactic globular cluster NGC6144 is located at $\alpha_{J2000}$=16$^h$27$^m$14.1$^s$ 
$\delta_{J2000}$=$-26^{\circ}01'29''$ (Alcaino. 1980) and E3 is located at  $\alpha_{J2000}$=09$^h$20$^m$59.3$^s$  
$\delta_{J2000}$=$-77^{\circ}16'57''$ (van den Bergh, Demers, and Kunkel 1980). Parameters of NGC6144 and E3 used in this paper are given in Table 1.

NGC6144 was observed for 55410s on 2007 July 6, 
and E3 was observed for 20110s on 2007 May 13, 
both with the Advanced CCD Imaging Spectrometer (ACIS) 
back-illuminated S3 chip on the \chandra\ X-ray Observatory. 
The field-of-view of the S3 chip ($\sim 8.3\times8.3$ arcmin$^2$) covers the entire half-mass radii of the two globular clusters. 
The data were collected with nominal frame time 3.2 sec.
We only focus on the sources detected on the S3 chip.

\subsection{Data Reduction and Sources Detection}

The data analysis was done with \chandra\ Interactive Analysis of Observations 
(CIAO) 3.4 software developed by the \chandra\ X-ray Center. 
We used CIAO wavelet-based \texttt{wavdetect}      
(Freeman et al. 2002) tool to detect discrete sources in 0.3-7 keV band. The light curve of the whole ACIS-S1 chip indicates there are no background flares. We also performed source detection  
on the 0.3-1, 1-2, and 2-7 keV images for both NGC6144 and E3. 
The detection threshold was set to be 10$^{-6}$ so that the false detection resulting from noise fluctuation is less than one from a $10^{6}$-pixel ACIS-S3 image. We set the wavelet radii from 1 to 16 pixels increasing by a factor of $\sqrt{2}$. 
A total of 36 X-ray sources for NGC6144 and 16 X-ray sources for E3 were detected. Among the sources detected by \chandra\ , 6 lie within the 
 half-mass radius($1'.62$) of NGC6144 while 3 lie within the half-mass radius($2'$.06) of E3 (Figure 1). The source counts were extracted from an elliptic region which covers 3$\sigma$  of the \chandra\ point spread function. The background count rate was extracted from the region without source outside the half-mass radii on the S3 chip.

Table 2 lists the detected \chandra\ X-ray sources in NGC6144 and E3. Sources were named by the total counts in 0.3-7keV band, and 6 sources out of NGC6144 (CX01 to CX06) as well as 3 sources out of E3 (CX01, CX02, CX03) lie in their own half-mass radius, respectively. The columns give the sources' name, the coordinates (J2000.0), the net counts of three energy bands which are soft (0.3-1keV), medium (1-2keV), hard (2-7keV), the 0.5-2.5 and 0.3-7keV counts and unabsorbed fluxes. The vignetting caused by off-axis flux from all sources within the half-mass radius accounts for less than 3\% of the change of flux. This value is less than the count error of each source. We therefore did not include vignetting calibration of flux of the source.
We use a power-law spectrum with a photon index of 1.5 and $N_H = 3.0 \times10^{21}$cm$^{-2}$ 
 for NGC6144 (estimating from Figure 2, see \S2.2), photon index of 2 and $N_H = 9.8 \times10^{20}$cm$^{-2}$ for E3 (Kalberla et al. 2005) to estimate the unabsorbed fluxes.

 We estimate the completeness limit to be $F_X = 1.62 \times10^{-15}$ergs s$^{-1}$cm$^{-2}$for NGC6144 and $F_X = 2.65 \times10^{-15}$ergs s$^{-1}$cm$^{-2}$ for E3 in 0.3-7keV energy band. This corresponds to $>$ 10 counts and $>$ 7 counts for NGC6144 and E3 anywhere on the S3 chip respectively.  
To obtain a rough estimate of the expected number of background sources expected to be found within the half-mass radius we proceed as follows. The area of the S3 chip that we analyze is $8.3'\times8.3'$. Writing the half-mass radius in arcminutes as $x$, and the number of sources detected outside the half-mass radius as $N$, we obtain the expected number of background sources within the half-mass radius as $N \pi x^2/(8.3^2-\pi x^2)$. This gives 4 and 3 background sources expected within the half-mass radii of NGC6144 and E3, respectively. The numbers of background sources estimated from the flux distribution in the Chandra Deep Field South (4 for NGC6144 and 2 for E3 respectively; Giacconi et al.\ 2001) are compatible with this. With 6 and 3 sources observed, was cannot exclude on the basis of the source numbers alone that all detected sources are due to background.  

\subsection{Count Rates, Light Curves and Spectral fitting}
We only extracted the energy spectra and light curves for sources with more than 100 counts: CX01, CX07, and CX08 in NGC6144. We rebin the spectra with 10 counts per spectral bin for CX01 and CX07, but 5 counts per spectral bin for CX08. Because of the low count numbers, we apply C-statistics in the spectral fitting (Nousek \& Shue 1989). We fit these three sources with simple absorbed power-law and thermal bremsstrahlung model(Table 3).The fitting of each source with both models seems equally well from the C-statistics. The light curves of CX01 and CX07 are 2ks per bin and CX08 3ks per bin. We did not find significant variation for any of the sources from their light curves.

The source counts are too few for the remaining sources to derive spectral parameters with meaningful results. Therefore, we also use hardness ratios to get rough X-ray spectral properties for all detected X-ray sources. These ratios were based on the source counts in three energy bands: S (0.3-1.0keV), M (1-2keV), and H (2-7keV). We analyzed two hardness ratios which are defined as HR1=(M-S)/(M+S) and HR2=(H-S)/(H+S). Figure 2 shows the X-ray color-color diagram for NGC6144 (left) and E3 (right) and is overlaid with four lines showing the tracks of different $N_H$ and photon index. From figure 2, a spectrum with photo index 1.5 and $N_H = 3.0 \times10^{21}$cm$^{-2}$ would give more accurate unabsorbed flux than that with $N_H = 1.28 \times10^{21}$cm$^{-2}$ given by Kalberla et al. (2005). The color-magnitude diagrams for all detected X-ray sources are displayed in Figure 3. 
  
\section{OPTICAL OBSERVATIONS AND DATA ANALYSIS}
Optical observations for NGC6144 and E3 were taken with the \hst\ Wide Field Camera of the Advanced Camera Surveys (ACS-WFC) with a field of view of $3.4'\times3.4'$, on 2006 April 15 (Proposal ID 10775). 
Both globular clusters were observed in F606W ($V_{606}$) and F814 ($I_{814}$) filters. 
For NGC6144, the exposure time with the F606W and F814W filters are 1725s and 1775s. For E3, the exposure time with F606W and F814W filters are both 405s. 
Two X-ray sources in NGC6144 (CX03 and CX05) are also included in an archival HST Wide Field and Planetary Camera 2 (WFPC2) observation (Proposal ID 11014). F336W ($U_{336}$), F439W ($B_{439}$), F675W ($R_{675}$), and F656N(H$\alpha$) filters were used in this observation. The exposure times were 3100s for F336W, 1700s for F439W, 3800s for F656N, and 900s for F675W filter respectively.
Figure 4 shows the ACS-WFC field of view covering the entire half-mass
radius of NGC6144 and most of the half-mass radius of E3, along with the field of view of the WFPC2 image for NGC6144 is also shown in Figure 4.

\subsection{Photometry}
\subsection*{ACS-WFC}
The photometry was done with the DOLPHOT package which is adapted from HSTphot (Dolphin 2000) for the use of ACS data.
Because of the resampling of images, 
the drizzled images produce suboptimal photometry; 
therefore we used the individual flat-fielded images for photometric reference and combined images for astrometry.
The \texttt{acsmask} masks out bad pixels from data quality images provided by the Space Telescope Science Institute (STScI),
and we got images in units of electrons on the raw images.
We employed \texttt{splitgroups} to split the ACS-WFC data with two chips into two extension files.  
The \texttt{calcsky} can create sky images for background determination. 
We use \texttt{acsfitdistort} to get the shift from the reference image to others.

There were five and four exposures with different physical offsets per filter 
for NGC6144 and E3, respectively. 
Two exposures with the same filter during the observations of 
NGC6144 and E3 have no relative offset. We therefore use them as reference images.
We gave the shift value for each chip, 
but the magnification and rotation are corrected by DOLPHOT. 
We ran photometry using DOLPHOT with ACS-WFC point-spread functions. 
DOLPHOT can analyse images within the same field and give individual photometry for each filter. Output photometry data were filtered to eliminate artifacts and obvious false detections.  

The photometries by Sarajedini et al. (2007) of NGC6144 are consistent with what we obtained by using DOLPHOT, but they do not provide magnitudes for all stars that are located in our error circles. However, in the photometry of E3 done by DOLPHOT, the stars were slightly distorted. The list from Sarajedini et al. provides better photometry for E3; we therefore use the result from Sarajedini et al. for the magnitudes of optical counterparts, whose color-magnitude diagrams (CMDs) are shown in Figure 5.

\subsection*{WFPC2}

Because the spatial resolution of ACS-WFC ($~0''.05$) is higher than WFPC2 ($~0''.1$ for WF and $~0''.05$ for PC), CX03a, CX03b, CX05a , and CX05b cannot be well resolved in WFPC2 observations. We therefore provide the source list from DOLPHOT and use HSTPHOT to do WFPC2 photometry. The position of stars in the source list was calibrated by \texttt{center} in IRAF. Some steps were performed before applying HSTPHOT. We use \texttt{mask} for data quality image provided by the STScI to mask bad pixel. We then remove the cosmic ray on the images by using \texttt{crmask}. The sky value is determined by \texttt{getsky}. The command \texttt{hotpixels} can remove all hot pixels. The measurements were done using a predetermined star list. The CMDs of WFPC2 F336, F656N, and F675 for NGC6144 are presented in Figure 6.    

\subsection{Astrometry}

In order to search for optical counterparts of \chandra\ X-ray sources in the fields, 
we have to improve the astrometry of each image. 
For NGC6144, we use relative astrometry instead of absolute astrometry 
to achieve a more accurate position. 
But for E3, the stellar density is relatively low. 
We used 2MASS (Two Micron All Sky Survey) point source catalog as reference
to improve the astrometry.

\subsubsection*{Astrometry for NGC6144}

For relative astrometry, we need to find sources observed both in the X-ray and optical bands, and determine the relative shift between the \hst\ ACS-WFC image and \chandra\ image. The optical stellar density of NGC6144 in the ACS-WFC field of view is too high to find reference stars; we can identify CX28 with a reliable optical source only, which would not be accurate enough to determine precise astrometry. Instead, we use an image of NGC6144 taken with a wide field imager. This image was taken by the European Southern Observatory (ESO) Max Planck Gesellschaft (MPI) 2.2m telescope with the Wide Field Imager (WFI) on 2002 February 21. The exposure time is 149.9s with the B-band filter. The images of the WFI were reduced using the Image Reduction and Analysis Facility (IRAF). The images were bias-subtracted and flat-fielded. To match coordinates on the wide field image with the HST image, we identify 23 isolated stars on the \hst\ image to overlay with the same set stars on the WFI image. This process was done by \texttt{ccmap} in IRAF with a registration error of $0''.06$. There are three isolated optical sources in the WFI corresponding to \chandra\ sources CX21, CX31 and CX32. The difference of coordinates on the \hst\ drizzled image to the \chandra\ image corresponds to a shift of $0''.156$ in R.A. and a shift of $0''.219$ in decl with an error of $0''.212$.

To search for possible optical counterparts, we combine three errors from astrometry and source detections to determine the radius of the error circle: the registration error between the \hst\ image and the WFI image, the error resulted from astrometry difference between \chandra\ and WFI, and the uncertainty of X-ray source positions detected by CIAO.
The final uncertainty on the position of X-ray sources is the quadratic sum of the above-mentioned uncertainties of the X-ray sources. The final astrometric solution gives an error of $0''.22$. The following identification will be focused on sources within the 95\% error circles.

\subsection*{Astrometry for E3}
To improve the absolute pointing accuracy of X-ray and optical frame, we use 2MASS All-Sky Catalog for Point Sources as reference for absolute position on both optical and X-ray images. We found 66 point sources on the \hst\ ACS-WFC image in a $3\times3$ arcmin$^{2}$  region around the center of E3 that matched entries in the 2MASS catalog.  
This gives positional rms residuals of $0''.173$ in R.A. and $0''.155$ in decl.
For the ACS-WFC image, the astrometric solution using the 66 stars yields rms residuals of $0''.134$ in R.A. and $0''.162$ in decl.
   
For the \chandra\ image, we found two X-ray sources (E3-CX01 and E3-CX14) match the 2MASS catalog. 
The calibration (using E3-CX01, E3-CX14 and the 2MASS counterparts) of coordinates on the \chandra\ image corresponds to a shift of $0''.134$ in R.A. and a shift of $0''.575$ in decl with residuals of $0''.062$ in R.A. and $0''.425$ in decl.

Combining the two 2MASS stars with rms residuals of $0''.183$ in R.A. and $0''.187$ in decl. gives the final astrometric solution with residuals of $0''.487$ in R.A. and $0''.560$ in decl.
For the astrometric calibration of X-ray and optical images, we get a $0''.805$ positional error. The error is the quadratic sum of the uncertainty of 2MASS sources for astrometry and the absolute astrometry for \hst\ and \chandra\ and the uncertainty of X-ray source position detected by CIAO. 

We later identify E3-CX01 in \chandra\ image with an HST object with certainty. This source in both images shares one same 2MASS star, so the uncertainty of the position of E3-CX01 is simply that given by \texttt{wavdetect}. 
However, while one optical source has been identified as a confirmed \hst\ counterpart to a \chandra\ source (E3-CX01), the error circles for other \chandra\ sources (E3-CX02 and E3 CX03) on the \hst\ image can be reduced to the sum of the \texttt{wavdetect} error for that source and the \texttt{wavdetect} error for the E3-CX01 (since relative HST positional errors are negligible, typically $< 0''.01$). Using the \texttt{wavdetect} errors in Table 2, we find the positional error for E3-CX02 is $0''.43$ and E3-CX03 $0''.38$. These errors are less than those estimated from absolute astrometry. Figure 7 gives sources within the 95\% error circles.

\section{SOURCES IDENTIFICATION}
To identify and classify sources in NGC6144 and E3, we analyse data from both X-ray and optical observations. 
We search for optical counterparts within the 95\% \chandra\ error circle. 
We discuss all X-ray sources within the ACS-WFC field of view but only consider sources within the half-mass radius to be cluster members in \S5 . For multiple \hst\ sources within the \chandra\ error circles, all the candidates in 95\% error circles are summarized in Table 4, Table 5 provide WFPC2 photometry for the candidates in the error circles of CX03 and CX05, and the finding chart is shown in Figure 7. 
Star just above the main sequence could be binaries in the cluster; stars left of the main-sequence could be binaries with accretion disks or hot white dwarfs. Other possibilities are foreground or background stars and background galaxies or AGNs.

The spectral characteristic is described by the hardness ratios (soft, medium and hard) in the X-ray color-color diagrams and color-magnitude diagrams.  
We also made a plot of the X-ray luminosity with respect to the optical V-band absolute magnitude (Figure 8). Based on previous studies on M4 (Bassa et al. 2004), NGC288 (Kong et al. 2006) and M55 \& NGC6366 (Bassa et al. 2008). The left diagonal line in Figure 8 roughly separates CVs and ABs. We calculated the X-ray flux to optical V-band apparent magnitudes as: log ($f_X$/$f_V$) = log $f_X$ + 5.67 +0.4$v_{606}$  (Green et al. 2004); $f_X$ is derived in the 0.5-2.5 keV band; we can roughly classify some sources with $f_X$/$f_V$ {$\geq$} 0.8 (Cool et al. 1995) and $M_V$ {$\leq$} 11 (Patterson 1984) as CVs.  

The \chandra\ X-ray sources NGC6144-CX28 and E3-CX03 are extended objects in the optical images and the optical color-magnitude diagrams also show that they are redder than main sequence stars, indicating they are background galaxies. 
The X-ray luminosity of NGC6144-CX01 ($L_{X(0.5-6keV)} \approx$ $1.8 \times 10^{32}$ erg $s^{-1}$) is too high for a binary (typically $L_{X(0.5-6keV)}<$ $10^{30}$ erg $s^{-1}$ from Verbunt \& Lewin 2006) of two main sequence stars; in addition, it has a high X-ray to optical luminosity ratio and a constant light curve. The optical counterpart to CX01 might be a chance coincidence with a background AGN. However, even in the case when we decrease the optical magnitude so the counterpart becomes fainter and CX01 shifts to the left in Figure 8, CX01 is still far from the parameter space occupied by galaxies in previous study. We suggest that NGC6144-CX01 is a possible CV, 
even though the optical counterpart on CMD lies on the main sequence. The spectrum of NGC6144-CX02 is relatively hard in X-rays. CX02 is bluer in optical and has a high X-ray to optical flux ratio. The optical magnitude suggests it is possibly a faint CV or a galaxy judging from its position in Figure 5 (similar to position in the CMD with M55-CX2 from Bassa et al. 2008).  
There are two optical counterparts in the error circle of NGC6144-CX03, CX03a and CX03b.  
The one possible optical counterpart (CX03a) is redder than the main sequence and the ratios of X-ray flux to optical V-band magnitudes is low ($<$0.2). However, CX03a is unlikely to be a point source, CX03a is a saturated star and may contain more stars. Certainly identification would not be possible without further observations. The other possible optical counterpart (CX03b) is bluer than the main sequence with $H\alpha$ emission, even the ratios of X-ray flux to optical V-band magnitudes is low ($<$0.5). As a result, we suggest that CX03 might be classified as a CV. 

The X-ray to optical flux ratio of NGC6144-CX04 is lower than CX03, and the optical counterparts (CX04a and CX04b) also lie on the main sequence; therefore it is also an possible AB. There are three possible optical counterparts in the error circle of the faint source NGC6144-CX05. The brightest counterpart (CX05c) saturated the \hst\ ACS-WFC detector; we can only analyse the other two counterparts.  
Two stars in the error circle of CX05, CX05a and CX05b, are on the main sequence. The X-ray to optical flux ratio for CX05a is 0.66; it may be an active binary. The X-ray to optical flux ratio of CX05b is somewhat high for an active binary. In the photometry of WFPC2 image, CX05a and CX05b are too faint (S/N $<$ 2) to provide reliable magnitudes. Classification of CX05b wouldn't be possible without further information. The lower limit (magnitude) of the saturated star (CX05c) in the error circle was 17.96 for $U_{336}$, 17.51 for $B_{439}$, 16.24 for $H\alpha$, 17.76 for $R_{656}$, 18.84 for $V_{606}$, and 15.63 for $I_{814}$. The I-band is three magnitudes brighter than V-band; hence it looks like a red giant. However, since it is saturated on the detector, we cannot assign it a correct color.  

For NGC6144-CX07, 
the most luminous X-ray source in NGC6144 has two possible optical counterparts:  CX07a and CX07b. The optical color-magnitude diagram of CX07a is bluer and CX07b redder, and the X-ray flux to optical flux ratio of CX07a and CX07b is 3.4 and 608.9 respectively. We identify CX07b as an AGN. Based on the optical results and X-ray hardness ratios of NGC6144-CX07, it could be either a CV or an AGN with CX07a as the optical counterpart. Certainly identification would not be possible without further observations.
We found two possible optical counterparts for NGC6144-CX08: CX08a and CX08b. Both of them appear on the main sequence, but the X-ray flux is still too high for CX08 to be an AB. The X-ray to optical flux ratio also indicates that 
NGC6144-CX08 is more likely to be a CV.

Based on the optical source density near the X-ray error circles the possibility of at least one the chance of at least one star falling by chance into the error circle of previous identification of ABs in this Section are: 35\% for CX03, 45\% for CX04, 91\% for CX05, and 44\% for CX08. We therefore cannot exclude the possibility that all our ABs are positional coincidence. 
Indeed, the upper limit for ABs (the dashed line in Figure 8) given by Bassa et al. (2008; Equation 3) indicates that CX04a is more likely to be a possible AB than other sources  higher than this limit. This limit for ABs is based on previous studies for stars in the solar neighborhood. We take this limit for AB in NGC6144 because the density of the cluster is low and the identified ABs in low core density globular cluster M12 (CX4b; Lu et al. 2009), M55 (CX7), and NGC6366 (CX4 and CX5; Bassa et al. 2008) are also lower than this limit. However, note that previous studies of ABs in high core density globular clusters can be higher than this limit. 
The position of CX04a in the CMD appears to be a subgiant, similar to the position of M55-CX7 in the CMD from Bassa et al. (2008). 
Some X-ray error circles contain more than one optical object without unusual features. 
They clearly require at least one optical object must be a coincidence.

The X-ray color-color diagram shows that E3-CX02 has a relatively hard X-ray spectrum compared to E3-CX01 and E3-CX03. The optical image of E3-CX02 also shows that it is a blue star and the high X-ray flux to optical magnitude suggests this is a possible CV. But the color of E3-02 is  bluer than turnoff. And it is one magnitude bluer than main sequence at same I-band magnitude. It appears bluer compared to CVs previously studied in globular clusters. Certainly identification would not be possible without further observations.

There are two unclassified X-ray sources (CX06 and CX30) in the ACS-WFC field of view of NGC6144 that no optical counterparts are found. The optical counterpart for E3-CX01 is too bright and saturates the CCD, it may be a foreground star. All other sources can not be classified without further observation. These two unidentified sources (CX06 and CX30) may be background sources which are too faint to be detected by \hst\ ACS-WFC in the W606F and W814F filters. 

We shift the error circle on the optical image of NGC6144 and E3 to eight different directions (respectively) about $5''$ for NGC6144 and $10''$ for E3 so that every time some sources would lie in the circle. According to the average number of sources found in shifting the circle (average over 8 trials in 8 directions), Table 6 listed the probabilities that the sources in each error circle come from positional coincidence. For CX04, based on the low probability ($<$6 \%) of finding two or more sources within the error circle, one of the two sources (CX04a, CX04b) is probably the real counterpart. For CX03 (the identified CV in this section), the probability that all the optical sources within the error circle are chance coincidence is also relatively low ($<$10\%). CX02 and CX05 are two marginal cases and we cannot exclude that all sources within the error circle are not real counterpart.      
Moreover, the probabilities that all sources in our error circle are coincidence (not true counterpart) are 0.06\% for NGC6144 and 0.92\%  for E3. Those low probabilities suggest that the potential candidates of optical counterparts in our error circle are very likely to be true counterparts.

\section{Discussion}
Based on the identification in \S4 and the probability of positional coincidence in Table 6, there is one good CV candidate (CX03) and two possible CVs (CX01 and CX02) within the half-mass radius of NGC6144. We also found one source (CX04) as active binary. The identification in \S4 indicates that at least 2 sources (CX03, CX04) are probable members associated with NGC6144 while other 2 sources (CX02, CX05) are marginal. We therefore assume 3 sources as cluster members of NGC6144 in the following discussion. If we take 3.5 sources to be chance coincidence in 6 error circles (from CX01 to CX06, see Table 6), the probability that at least 3 sources are cluster members is higher than 86\%. Three sources in globular cluster E3 were detected within the half-mass radius, with one insecure CV. The study by Pooley et al. (2003) indicates the relation between the stellar encounter rate and the number of X-ray sources with 
$L_{X(0.5-6keV)}>$  $4\times10^{30}$ ergs s$^{-1}$ in globular clusters. The encounter rates are proportional to $\rho_c{^2}$$r_c^{3}$$\upsilon_{rms}^{-1}$ where $\rho_c$ is the central density of globular cluster, $r_c$ the core radius, and $\upsilon_{rms}$ the central velocity dispersion. We estimate the encounter rate by $\Gamma$$\varpropto$ $\rho_c^{1.5}$$r_c^{2}$ via $\upsilon_{rms}$ $\varpropto$ ${(\rho_cr_c)}^{0.5}$ (Verbunt 2003).  
The encounter rate for NGC6144 ($\Gamma_{6144}=0.07$) is about 1/330 of 47 Tuc and 1/15 of M4. For E3 ($\Gamma_{E3}=0.001$) it is about 1/15000 of 47 Tuc and 1/700 of M4. The relevant detected numbers of X-ray sources (
$L_{X(0.5-6keV)}>$ $4\times10^{30}$ ergs s$^{-1}$ ) in 47 Tuc and M4 are 45 and 6 respectively (Pooley et al. 2003).
It is clear that the X-ray sources number in low core density globular cluster, NGC6144 (2-3 sources) and E3 (one doubtful source), does not scale with $\Gamma$.

We consider the number of X-ray sources with the collision number and mass (Pooley \& Hut 2006).  
The best fit for a predicted number of X-ray sources in a globular cluster by Bassa et al. (2008) is $N=1.2\Gamma+1.1M_h$. Values for $\Gamma$ and $M_h$ are normalized to the value of M4. According to this fitting, one X-ray source is predicted in NGC6144. 

We also repeated the fitting based on Bassa et al.(2008) with the same clusters (Pooley et al. 2003) as well as NGC288($\mu_b$=8), M55($\mu_b$=7), NGC6366($\mu_b$=4),  NGC6144($\mu_b$=4.5), and NGC6218($\mu_b$=4; Lu et al. 2009) where $\mu_b$ is the expected number of background sources (Verbunt et al. 2007).
Verbunt et al.(2007) suggest a Poisson distribution to overcome the small-number statistics of clusters with small source numbers.
According to the Poisson distribution, the probability of observing N X-ray sources when $\mu$ is expected is P(N,$\mu$)=$\mu^N$ $e^{-\mu}$/N!. The fitting procedure consists of varying a and b in $N = a\Gamma + bM_h$ to maximize the probability 
P=$\prod_{j}$[P($N_c$,$\mu_c$)P($N_b$,$\mu_b$)]$_j$ 
,where j indexes the clusters, $N_b$ the number of background sources not related to the cluster, $N_c$ the number of cluster sources within the half-mass radius (subtract $N_b$ from observed number of X-ray sources), and $\mu_c$ the expected number of cluster sources. 
For these fittings we assumed the lower limit of source number to be 2 for NGC288, 3 for M55, 2 for NGC6366, 2 for NGC6218 (Lu et al. 2009) and 3 for NGC6144. The lower limit of the other clusters sources is 0.
The best solution is found for $N=1.23\Gamma+1.14M_h$ with the total probability P=$e^{-57}$ (see Verbunt et al. 2007). This result is consistent with Bassa et al. (2008) even when we include NGC6144 and NGC6218. Figure 9 gives the comparison between the observed number of sources (background-subtracted) and the best fit number.

We fitted the data anew, excluding NGC6218, NGC288, M55 and NGC6366. The best solution is found to be $N=1.41\Gamma+0.62M_h$ with P=$e^{-37}$(without lower limit to the source number of NGC6144) or $N=1.32\Gamma+0.83M_h$ with P=$e^{-38}$ (with lower limit to the source number of NGC6144 equaling to 3). 
According to the previous fits result $N=1.23\Gamma+1.14M_h$, 
the first term, which is the stellar encounter rate, yields 0.1 expected sources ($\Gamma_{6144}=0.07$), the second term, which is the mass of NGC6144, yields 1.0 expected source ($M_{6144}=0.87$). 
The dependence of the X-ray sources in NGC6144 on mass is significantly shown.

If E3-CX02 is the member of E3, then the number of identified X-ray sources in globular cluster E3 is still higher than the prediction based on the relation between the mass of the cluster and the number of sources. This may be due to the structure of the cluster is now being disrupted under the galactic tidal force (van den Bergh et al. 1980). The encounter rate and the mass of E3 are 0.001 and 0.022 (normalized with M4), respectively. The X-ray sources in E3 should have more dependence on mass than $\Gamma$. 

The scaling of the number of detected source number found with $L_{X(0.5-6keV)}>$  $4\times10^{30}$ ergs s$^{-1}$ (Pooley et al. 2003; see also Verbunt et al. 2002; Heinke et al. 2003) suggests that CVs are mostly made from close encounters. The population of CVs may be caused by either tidal capture or interactions involving primordial binaries (Davies 1997). 
In a low core density globular cluster, primordial CV-progenitor binaries are likely to survive long enough to form CVs in the cluster core. NGC6144 and E3 are the two clusters selected to study the faint X-ray sources population in low core density globular clusters.
By combining both the \chandra\ and \hst\ data, we found possible CVs in NGC 6144, and no good CV candidates in E3.
Since these CVs are unlikely to be formed via dynamical processes (stellar encounter), they must be evolved from primordial binaries as predicted by Davies (1997).

There is evidence that the number of X-ray sources in low core density globular cluster does not scale with $\Gamma$ are also shown in the globular cluster NGC288 (also Kong et al. 2006). The study of M55 and NGC6366 by Bassa et al. (2008) clarifies the dependence of the number of X-ray sources not only on $\Gamma$ but also on the mass of the globular cluster. In brief, there are evidences that close binaries in globular cluster could be primordial in origin.

\begin{acknowledgements}
We thank William E. Harris for supplying catalog of parameters in our galaxy.
We are grateful to Andrew Dolphin for his DOLPHOT code and some assistance to solve photometric problems. Special thanks to Yen Tzu-Ching for her help in revising the grammar and coherence of this paper. This research is based on observations made with the NASA/ESA Hubble Space Telescope, obtained from the Data Archive at the Space Telescope
Science Institute, which is operated by the Association of
Universities for Research in Astronomy, Inc., under NASA contract NAS5-26555. These observations are associated with \hst\ proposal number 10775 and 11014. This project is supported by the National Science Council of Republic of China, Taiwan through grant NSC 96-2112-M-007-037-MY3.
 
\end{acknowledgements}

\begin{figure*}
\medskip
\psfig{file=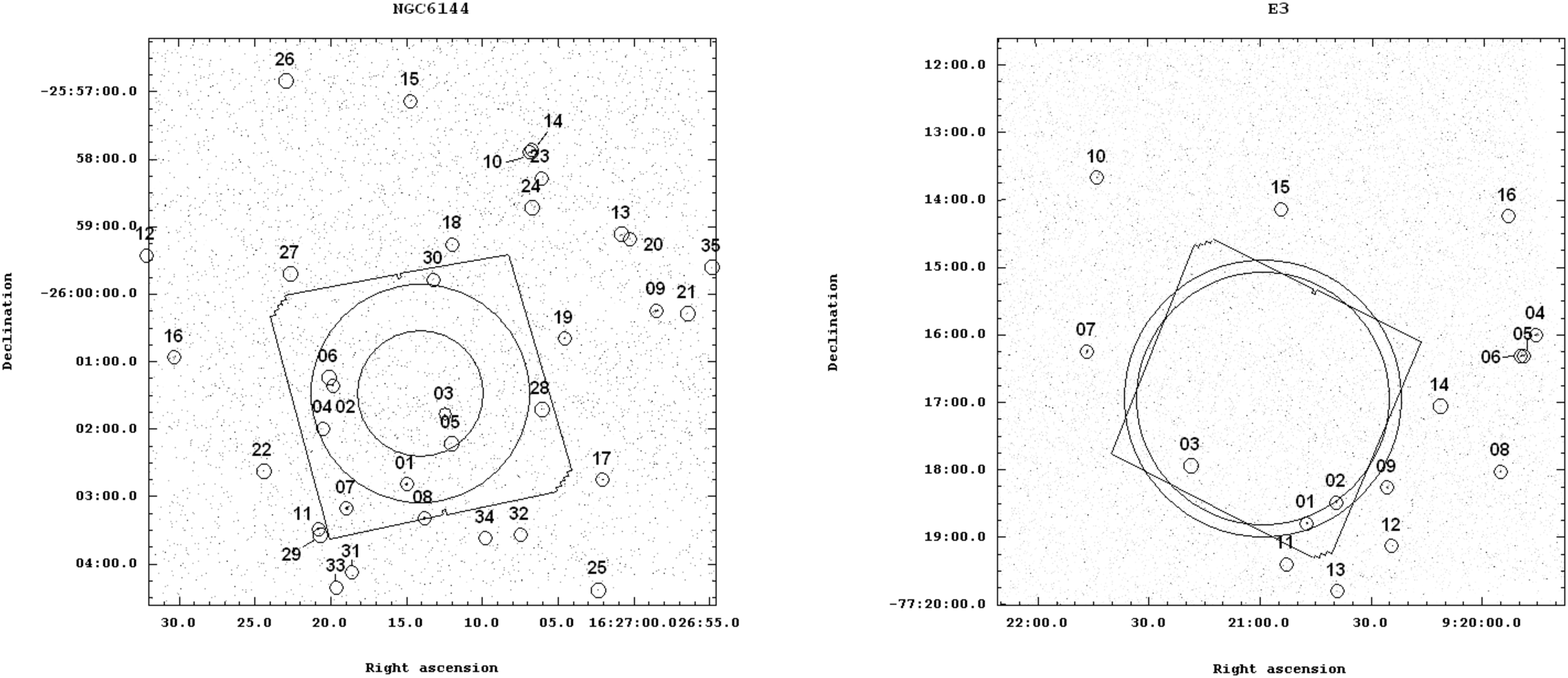,width=6.0in}
\caption{\chandra\ X-ray sources on the entire ACIS-S3 chips with 0.3-7 keV 
of NGC6144 (left) and E3 (right). 
The square is the field of view of \hst\ ACS-WFC. 
The large circle is the half-mass radius of the cluster 
and the smaller circle the core.
All detected X-ray sources are marked with source number.}
\end{figure*}

\vspace{2mm}
\begin{figure*}
\psfig{file=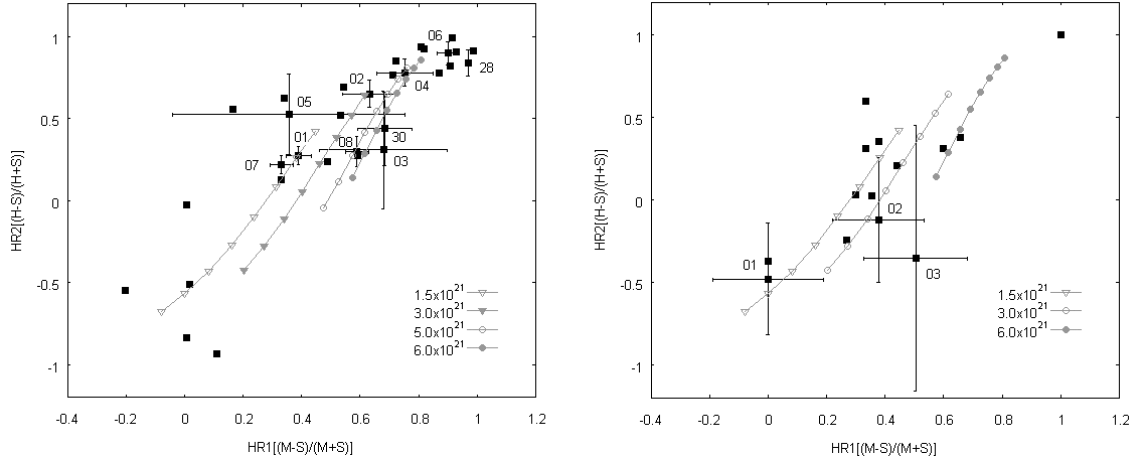,width=6.0in}
\caption{X-ray color-color diagram for most of the X-ray sources detected in \chandra\ S3 chip of NGC6144 (left) and E3 (right). The sources marked with error bars and source names are the ones that lie in the field of view of \hst\ ACS-WFC. We fit the data with power law model. The values of $N_H$ in NGC6144 are $1.5\times10^{21}$, $3.0\times10^{21}$, $5.0\times10^{21}$, and $6.0\times10^{21}$ cm$^{-2}$ (from left to right). We also applied power-low 
model on E3; the values of $N_H$ are $1.5\times10^{21}$, $3.0\times10^{21}$, and $6.0\times10^{21}$(from left to right). For each power-law model, the photon index varies as (from bottom to top) $\alpha$=3.0, 2.75, 2.5, 2.25, 2.0, 1.75, 1.5, 1.25.}

\end{figure*}

\vspace{2mm}
\begin{figure*}
\psfig{file=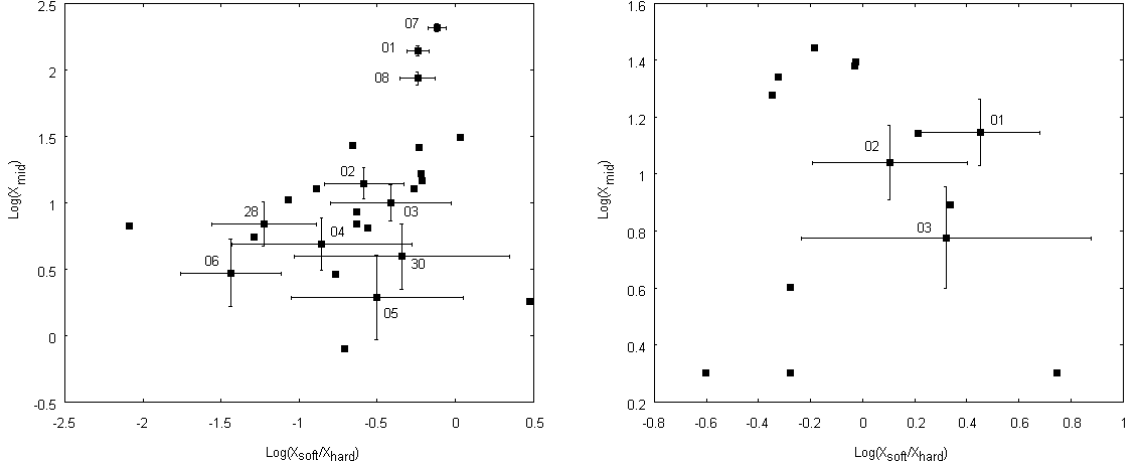,width=6.0in}
\caption{X-ray color-magnitude diagram for most of the X-ray sources detected in \chandra\ S3 chip of NGC6144 (left) and E3 (right). The X-ray color is
  defined as the logarithm of the ratio of 0.3--1.0 keV (X$_{soft}$) counts to
  2.0--7 keV ($X_{hard}$) counts, and the magnitude is the logarithm of 1.0--2.0 keV
  ($X_{mid}$) counts. The sources marked with error bars are the ones that lie in the field of view of \hst\ ACS-WFC.}
\end{figure*}

\vspace{2mm}
\begin{figure*}
\psfig{file=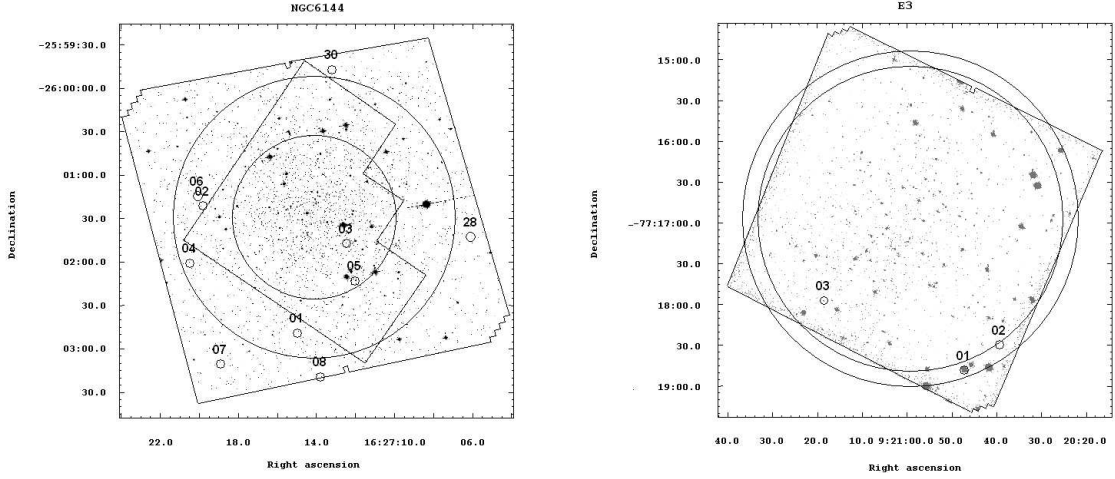,width=6.0in}
\caption{The \hst\ ACS-WFC images with X-ray sources detected by \chandra\ with 0.3-7 keV of NGC6144 (left) and E3 (right). The square is the field of view of \hst\ ACS-WFC. The polygon on NGC6144 is the field of view of \hst\ WFPC2. Half-mass and core radii are shown.}
\end{figure*}

\vspace{2mm}
\begin{figure*}
\psfig{file=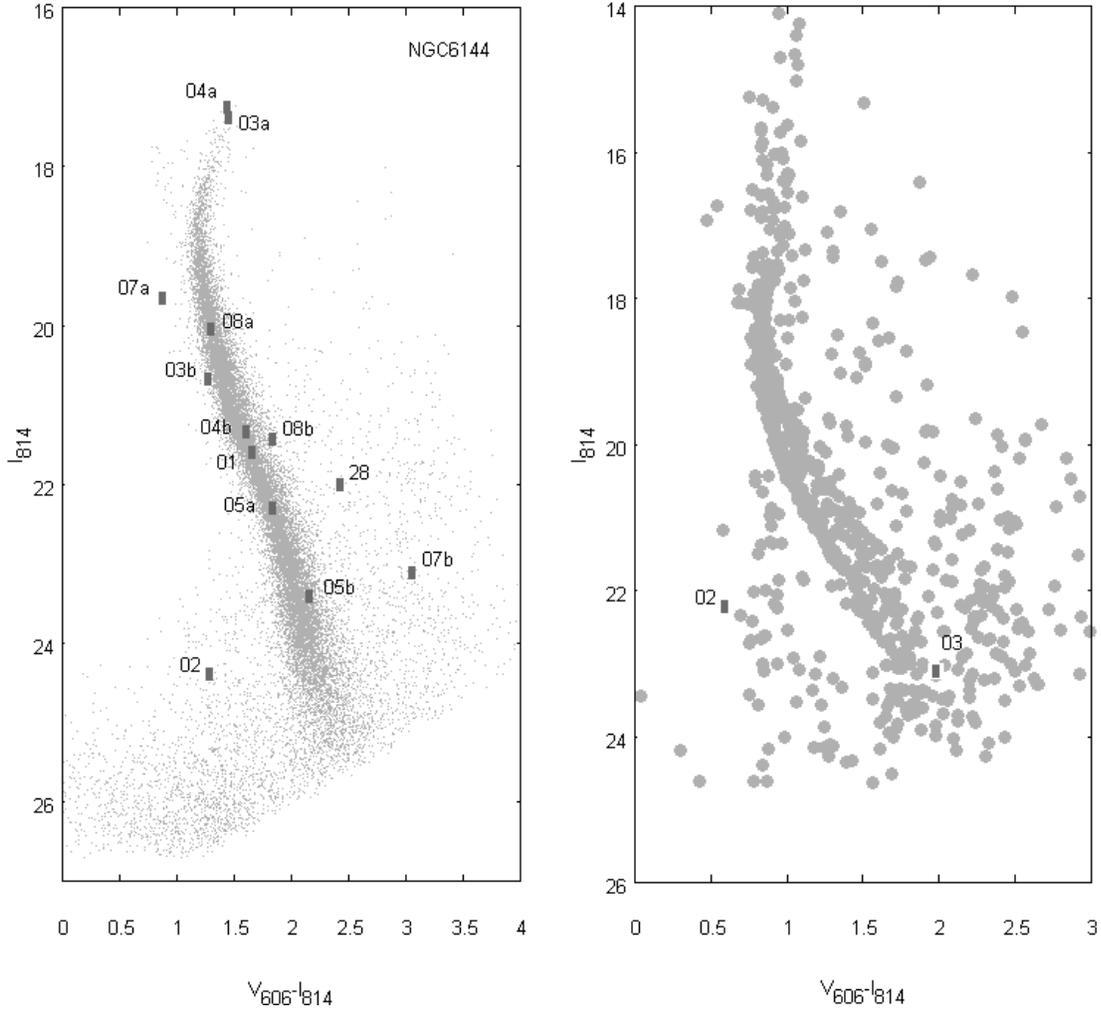,width=6.0in}
\caption{Optical color-magnitude of the \hst\ ACS-WFC observation of NGC6144(left) and E3(right). Possible X-ray source candidates for optical counterparts are labeled with source name. The letters after source names refer to the candidate optical counterparts to the X-ray sources. }
\end{figure*}

\vspace{2mm}
\begin{figure*}
\psfig{file=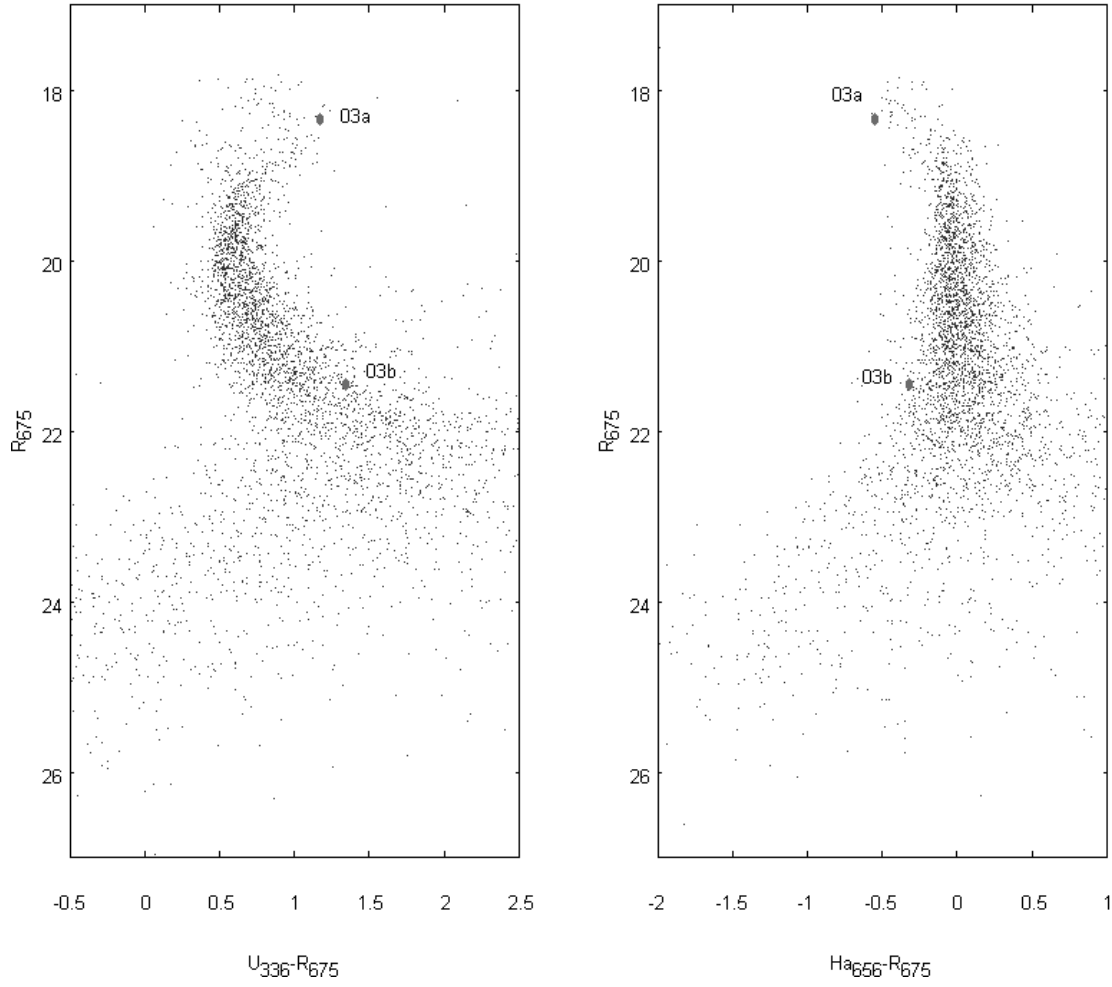,width=6.0in}
\caption{Optical color-magnitude of the \hst\ ACS-WFPC2 observation of NGC6144. Only CX03a and CX03b provide reliable magnitudes. }
\end{figure*}

\vspace{2mm}
\begin{figure*}
\psfig{file=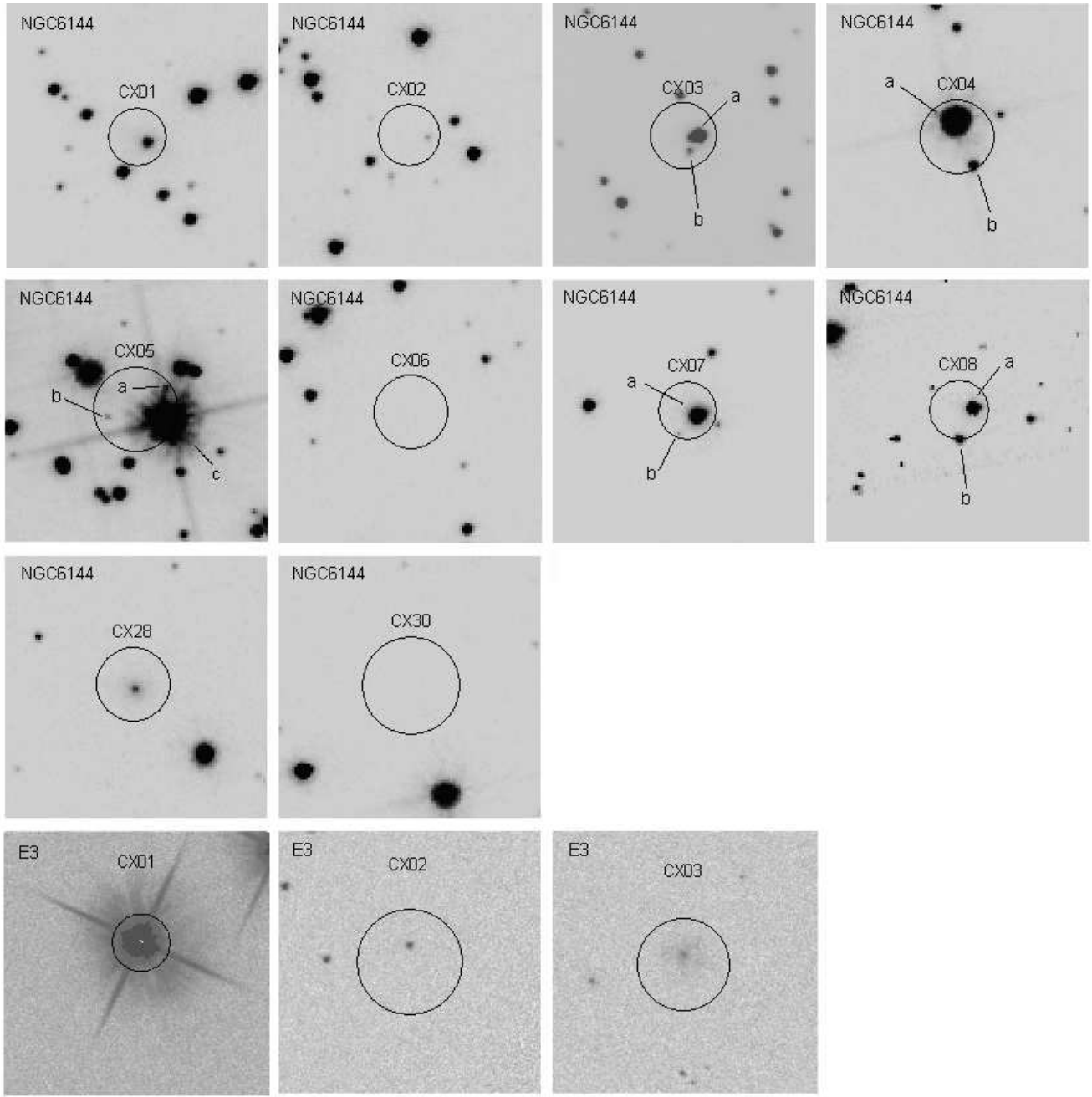,width=7.0in}
\caption{The ($5\times5$ arcsec$^{2}$) view the candidate optical counterparts of NGC6144 and ($10\times10$ arcsec$^{2}$) view of E3. These images were taken from \hst\ ACS-WFC observations in F606W. The 95\% error circles were overlaid for the \chandra\ source position. All images of NGC6144 use the same gray scale throughout except for CX03. The images of E3 use the same gray scale throughout but different from that used for NGC6144.}
\end{figure*}

\vspace{2mm}
\begin{figure*}
\psfig{file=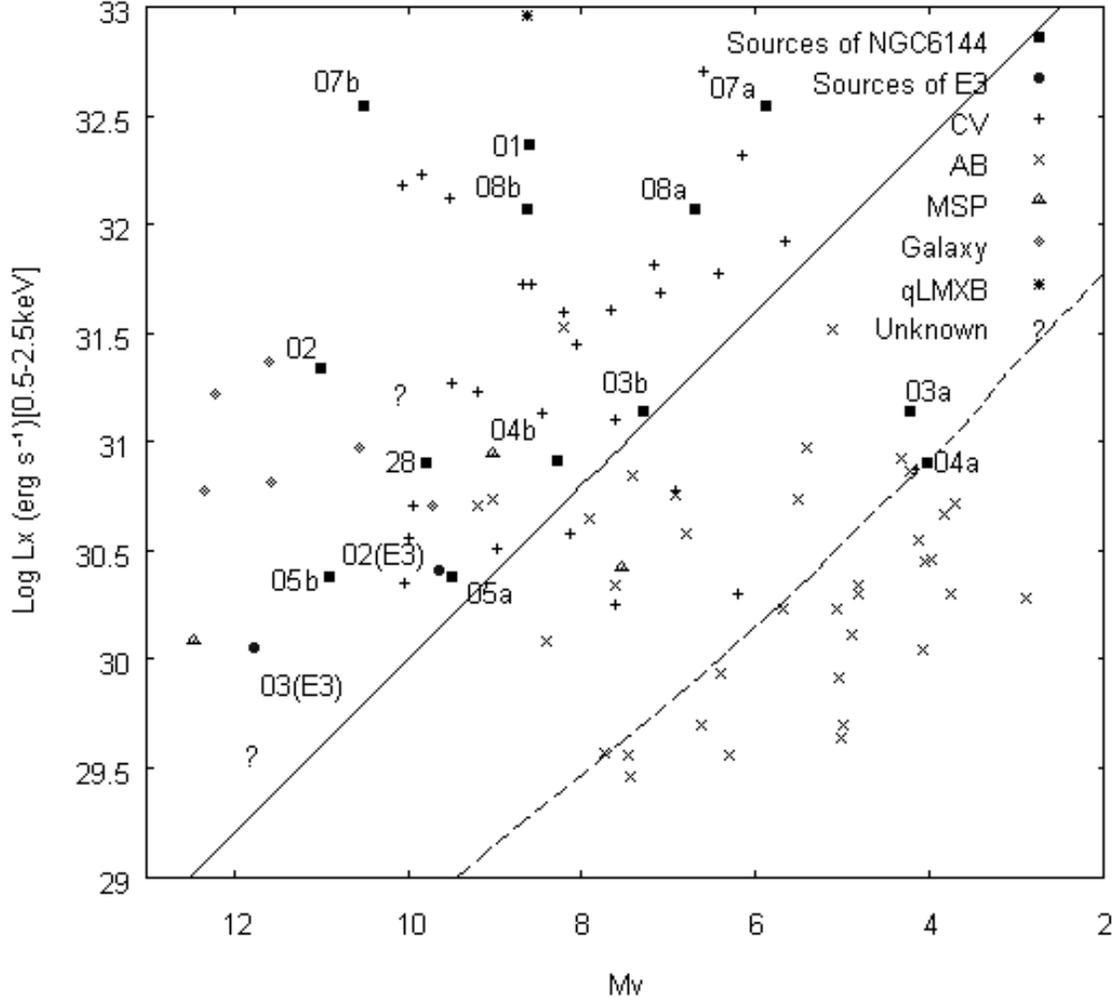,width=6.0in}
\caption{X-ray luminosity to V-band absolute magnitude diagram. The X-ray luminosity $L_x$ is in the range from 0.5keV to 2.5keV. The classified X-ray sources from NGC288 , 47Tuc, NGC6121, and NGC6752 are also plotted with NGC6144 and E3. Optical magnitude of V band for NGC6144 and E3 are observed using \hst\ F606W filter. 
The solid line was used in earlier papers (e.g. Bassa et al. 2004) to separate cataclysmic variables above it from
active binaries below it. However, the maximum X-ray luminosity
observed for active binaries near the Sun is about 0.001
$L_\mathrm{bol}$ (Verbunt et al . 2007, Bassa et al.  2008),
and this line is indicated with a dashed line (based on
$L_{bol}$ from the 11.2 Gyr isochrone for metal-poor z=0.001
stars as computed by Girardi et al. 2000).}

\end{figure*}

\vspace{2mm}
\begin{figure*}
\psfig{file=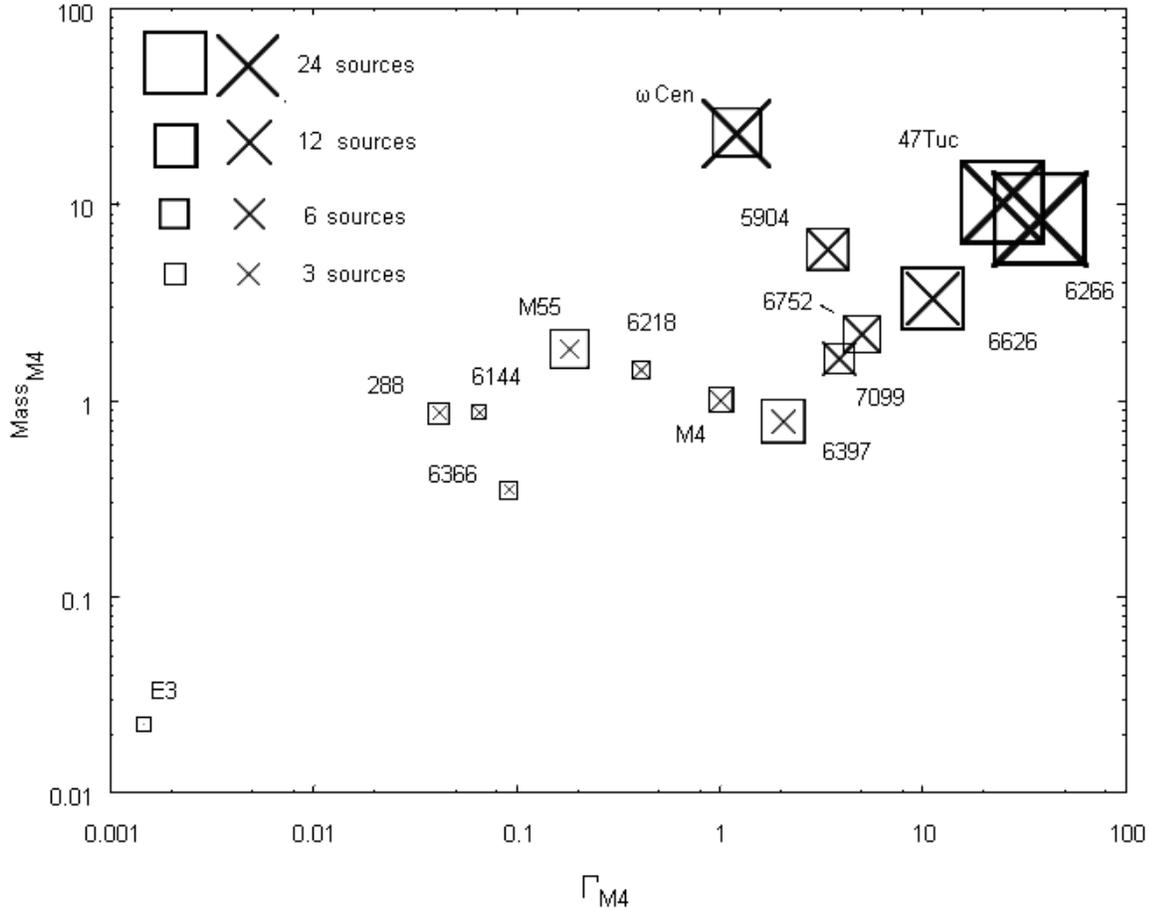,width=6.0in}
\caption{Comparison between the observed number of sources (symbols: $\square$) and the best fit number (symbols: $\times$). The position of each cluster is given by $\Gamma$ and Mass. The scales for the number of sources are on the top left of the figure.
The best fit gives number of each cluster by $N=1.2\Gamma+1.1M_h$. The size of the symbols is proportional to their number. The mass and encounter number($\Gamma$) of clusters are normalized with M4(NGC6121).}
\end{figure*}


\begin{table*}
\caption{Parameters of NGC6144 and E3}
\begin{tabular}{lcccccc}
\hline \hline
Cluster  & $r_c$ & $r_h$ & d (kpc) & log$\rho_c$ &$E_{B-V}$ & $N_H$ cm$^{-2}$\tablenotemark{*}\\
\hline
NGC6144 & $0.94'$& $1.62'$ & 8.5 & 2.23 & 0.36 & $1.28\times10^{21}$\\
E3 & $1.87'$ & $2.06'$ & 4.3 & 1.11 & 0.3 & $9.36\times10^{20}$\\

\hline
\end{tabular}
\par
\medskip
\begin{minipage}{0.82\linewidth}
NOTES. --- Values of angular core radius $r_c$, angular half-mass radius $r_h$, distance d, luminosity densities $\rho_c$, reddening $E_{B-V}$, Neutral hydrogen column density $N_H$ .\\
$^*$ The interstellar column density of neutral hydrogen was taken by the average 21-cm emission from 1 degree$^{2}$ around the center of the globular cluster taken from The Leiden/Argentine/Bonn (LAB) Survey of Galactic HI. \\
\end{minipage}
\end{table*}

\begin{table*}
\centering{\scriptsize
\caption{\chandra\ Source Properties }
\begin{tabular}{lcccccccc}
\hline
\hline
\multicolumn{2}{c}{Source} &\multicolumn{5}{c}{Net Counts} & \multicolumn{2}{c}{Flux}  \\
\cline{1-2} \cline{3-7} \cline{8-9}

Source Name & R.A.  Decl. (J2000.0) & Soft & Medium & Hard &$0.5-2.5$& $0.3-7$ & $F_{0.5-2.5}$& $F_{0.3-7}$ 
\\
\hline
\\
\multicolumn{9}{c}{NGC6144}\\

CX01 	&	 16:27:14.97(0.02) -26:02:49.13 (0.02)	&	62.8	&	138.9	&	108.8	&	228.8	&	310	&	29.15 	&	64.33 	\\
CX02 	&	 16:27:19.83(0.09) -26:01:21.22 (0.08)	&	3.9	&	13.9	&	14.8	&	17.8	&	32.6	&	2.27 	&	6.76 	\\
CX03 	&	 16:27:12.47(0.11) -26:01:47.17 (0.09)	&	1.9	&	9.9	&	4.7	&	13.8	&	16.6	&	1.76 	&	3.44 	\\
CX04 	&	 16:27:20.48(0.10) -26:02:00.73 (0.12)	&	0.8	&	4.9	&	5.8	&	7.7	&	11.4	&	0.98 	&	2.36 	\\
CX05 	&	 16:27:12.02(0.20) -26:02:13.09 (0.11)	&	0.9	&	1.9	&	2.9	&	2.8	&	5.7	&	0.35 	&	1.19 	\\
CX06 	&	 16:27:20.10(0.18) -26:01:14.33 (0.14)	&	0.1	&	3	&	1.9	&	3.9	&	4.8	&	0.50 	&	1.00 	\\
\\
CX07 	&	 16:27:18.93(0.02) -26:03:10.52 (0.02)	&	103.8	&	206.8	&	133.6	&	341.6	&	444.2	&	43.48 	&	92.23 	\\
CX08 	&	 16:27:13.81(0.04) -26:03:19.34 (0.03)	&	21.8	&	84.9	&	39.7	&	112.7	&	146.4	&	14.33 	&	30.36 	\\
CX09 	&	 16:26:58.50(0.10) -26:00:14.79 (0.08)	&	0.6	&	6.7	&	73.3	&	11.3	&	80.5	&	1.44 	&	16.74 	\\
CX10 	&	 16:27:06.89(0.20) -25:57:53.77 (0.14)	&	11.6	&	31	&	10.8	&	45.9	&	53.5	&	5.84 	&	11.11 	\\
CX11 	&	 16:27:20.80(0.07) -26:03:29.34 (0.08)	&	9.7	&	25.8	&	16.5	&	40.5	&	52	&	5.16 	&	10.80 	\\
CX12 	&	 16:27:32.16(0.21) -25:59:25.78 (0.25)	&	4	&	26.8	&	18.2	&	36.4	&	49	&	4.62 	&	10.18 	\\
CX13 	&	 16:27:00.80(0.14) -25:59:06.63 (0.16)	&	2.1	&	10.4	&	24.6	&	17.7	&	37.2	&	2.24 	&	7.71 	\\
CX14 	&	 16:27:06.71(0.19) -25:57:51.94 (0.16)	&	5.3	&	16.5	&	8.8	&	26.9	&	30.5	&	3.41 	&	6.33 	\\
CX15 	&	 16:27:14.75(0.31) -25:57:08.42 (0.30)	&	4.8	&	14.7	&	7.8	&	20	&	27.3	&	2.55 	&	5.66 	\\
CX16 	&	 16:27:30.28(0.14) -26:00:56.19 (0.13)	&	1.6	&	12.7	&	12.4	&	16.4	&	26.7	&	2.08 	&	5.55 	\\
CX17 	&	 16:27:02.05(0.14) -26:02:45.02 (0.14)	&	4.6	&	12.7	&	8.4	&	17.4	&	25.7	&	2.21 	&	5.35 	\\
CX18 	&	 16:27:11.97(0.15) -25:59:15.99 (0.12)	&	0	&	5.7	&	18.3	&	9.3	&	23.5	&	1.19 	&	4.89 	\\
CX19 	&	 16:27:04.57(0.14) -26:00:39.50 (0.12)	&	0	&	12.7	&	10.4	&	15.4	&	22.7	&	1.97 	&	4.71 	\\
CX20 	&	 16:27:00.25(0.24) -25:59:10.71 (0.22)	&	3.2	&	6.4	&	11.7	&	7.7	&	21.2	&	0.98 	&	4.41 	\\
CX21 	&	 16:26:56.44(0.32) -26:00:17.39 (0.24)	&	8.8	&	11.4	&	0	&	21.7	&	19.8	&	2.75 	&	4.12 	\\
CX22 	&	 16:27:24.43(0.10) -26:02:37.55 (0.11)	&	1.8	&	6.9	&	7.7	&	9.7	&	16.4	&	1.23 	&	3.42 	\\
CX23 	&	 16:27:06.11(0.26) -25:58:16.94 (0.21)	&	1.4	&	8.5	&	6	&	12	&	15.9	&	1.53 	&	3.29 	\\
CX24 	&	 16:27:06.70(0.28) -25:58:43.15 (0.19)	&	0	&	8.3	&	7.5	&	9.6	&	14.9	&	1.22 	&	3.11 	\\
CX25 	&	 16:27:02.35(0.16) -26:04:23.01 (0.10)	&	1.8	&	2.9	&	9.7	&	5.7	&	14.4	&	0.73 	&	2.98 	\\
CX26 	&	 16:27:22.93(0.26) -25:56:50.40 (0.20)	&	0.3	&	5.5	&	5.8	&	8.9	&	11.6	&	1.13 	&	2.39 	\\
CX27 	&	 16:27:22.65(0.18) -25:59:41.99 (0.31)	&	0.6	&	5.7	&	4.3	&	9.3	&	10.5	&	1.19 	&	2.19 	\\
CX28 	&	 16:27:06.09(0.16) -26:01:42.59 (0.08)	&	0.1	&	6.9	&	1.8	&	7.8	&	8.7	&	1.00 	&	1.81 	\\
CX29 	&	 16:27:20.68(0.16) -26:03:33.78 (0.18)	&	0.8	&	2.9	&	4.7	&	3.7	&	8.4	&	0.47 	&	1.74 	\\
CX30 	&	 16:27:13.21(0.19) -25:59:47.07 (0.15)	&	0.9	&	4.8	&	1.7	&	6.6	&	7.3	&	0.84 	&	1.51 	\\
CX31 	&	 16:27:18.57(0.20) -26:04:07.40 (0.14)	&	3.8	&	3.8	&	0	&	7.6	&	7.2	&	0.97 	&	1.50 	\\
CX32 	&	 16:27:07.47(0.13) -26:03:34.23 (0.10)	&	2.9	&	3	&	0	&	5.9	&	5.8	&	0.75 	&	1.20 	\\
CX33 	&	 16:27:19.61(0.25) -26:04:20.80 (0.14)	&	0.9	&	1.9	&	2.8	&	2.8	&	5.6	&	0.35 	&	1.16 	\\
CX34 	&	 16:27:09.79(0.21) -26:03:37.26 (0.19)	&	0.7	&	0.8	&	3.6	&	1.6	&	5.1	&	0.21 	&	1.07 	\\
CX35 	&	 16:26:54.85(0.27) -25:59:35.97 (0.17)	&	1.8	&	1.8	&	0.6	&	1.6	&	4.2	&	0.21 	&	0.88 	\\
CX36 	&	 16:26:56.74(0.61) -25:56:04.29 (0.40)	&	1	&	0	&	2.2	&	0	&	3.1	&	0.21 	&	0.63 	\\

\hline

\\
\multicolumn{9}{c}{E3}\\

CX01 	&	 09:20:47.31(0.17) -77:18:48.46 (0.15)	&	14	&	13	&	4.9	&	29	&	31.9	&	7.12 	&	12.30 	\\
CX02 	&	 09:20:39.36(0.28) -77:18:29.72 (0.23)	&	5	&	11	&	3.9	&	15.9	&	19.8	&	3.92 	&	7.66 	\\
CX03 	&	 09:21:18.57(0.22) -77:17:57.16 (0.21)	&	2	&	6	&	1	&	7	&	8.9	&	1.71 	&	3.43 	\\
\\
CX04 	&	 09:19:45.67(0.32) -77:16:00.30 (0.41)	&	10.8	&	27.8	&	16.5	&	44.6	&	55.1	&	10.96 	&	21.26 	\\
CX05 	&	 09:19:48.94(0.41) -77:16:18.89 (0.17)	&	9.9	&	21.9	&	20.8	&	35.8	&	52.6	&	8.81 	&	20.28 	\\
CX06 	&	 09:19:49.59(0.27) -77:16:19.01 (0.18)	&	12.9	&	23.9	&	13.8	&	39.9	&	50.7	&	9.79 	&	19.50 	\\
CX07 	&	 09:21:46.49(0.35) -77:16:15.19 (0.26)	&	11.8	&	24.8	&	12.5	&	42.7	&	49.2	&	10.47 	&	18.99 	\\
CX08 	&	 09:19:55.11(0.42) -77:18:01.89 (0.39)	&	3.9	&	18.9	&	8.7	&	22.8	&	31.6	&	5.60 	&	12.19 	\\
CX09 	&	 09:20:25.74(0.24) -77:18:16.33 (0.19)	&	8	&	13.9	&	4.9	&	21.9	&	26.8	&	5.38 	&	10.34 	\\
CX10 	&	 09:21:43.68(0.58) -77:13:40.80 (0.42)	&	7.8	&	7.8	&	3.6	&	15.7	&	19.3	&	3.86 	&	7.43 	\\
CX11 	&	 09:20:52.90(0.22) -77:19:24.57 (0.19)	&	5	&	2	&	0.9	&	7	&	7.9	&	1.71 	&	3.04 	\\
CX12 	&	 09:20:24.39(0.36) -77:19:08.32 (0.20)	&	1	&	2	&	4	&	4	&	6.9	&	0.98 	&	2.67 	\\
CX13 	&	 09:20:39.19(0.33) -77:19:48.60 (0.24)	&	1	&	4	&	1.9	&	5	&	6.9	&	1.22 	&	2.65 	\\
CX14 	&	 09:20:11.36(0.56) -77:17:03.80 (0.39)	&	0	&	4	&	2.9	&	4.9	&	6.9	&	1.21 	&	2.65 	\\
CX15 	&	 09:20:54.27(0.67) -77:14:09.02 (0.54)	&	1	&	2	&	1.9	&	2.9	&	4.9	&	0.72 	&	1.88 	\\
CX16 	&	 09:19:53.44(0.31) -77:14:14.59 (0.29)	&	1	&	1	&	2	&	3	&	4	& 0.73 	&	1.53 	\\

\hline

\end{tabular}

\par
\medskip
\begin{minipage}{0.95\linewidth}
NOTES. --- 
List of \chandra\ X-ray sources in our observations of NGC6144 (upper part) and E3(bottom part). The positions of stars were given by event-2 file from \chandra\ observations and calibrated by using Two Micron All Sky Survey (2MASS) of Point Sources as reference. 
The position errors  given by \texttt{wavdetect} are listed between brackets, in arcsec. The signal to noise ratios of photon counts were around 2-21. 
The source counts have been corrected for background. The first 6 sources  are located within the half-mass radius of NGC6144 and are numbered according to total counts in 0.3-7keV band. The first 3 sources  are located within the half-mass radius of E3 and are numbered according to total counts in 0.3-7keV band. The remaining sources are located outside the half-mass radius and are numbered according to total count in 0.3-7keV band. The unabsorbed flux is in units of $10^{-15}$ ergs $cm^{-2}$ $s^{-1}$.
\end{minipage}
}
\end{table*} 

\begin{table*}
\caption{Spectral Fits of the Brightest Sources}
\begin{tabular}{lccccc}
\hline \hline
Source Name & Model\tablenotemark{a} & $N_H$\tablenotemark{b} & $kT/\alpha$ & $Goodness$\tablenotemark{c} & $f_{0.3-7}$\tablenotemark{d}\\
\hline
CX01 & PL & $0.78_{-0.70}^{+1.22}$& 1.23$\sim$1.63 & 73\%& 5.2\\
  & TB & $0.38_{-0.31}^{+1.63}$& $<$ 49.84 & 79\%& 4.5\\
CX07 & PL & $2.34_{-0.53}^{+0.75}$& 1.70$\sim$2.20 & 15\%& 8.3\\
  & TB & $1.51_{-0.39}^{+0.51}$& 4.04$\sim$7.74 & 10\%& 4.4\\
CX08 & PL & $4.16_{-0.34}^{+2.27}$& 1.79$\sim$2.35 & 78\%& 3.7\\
  & TB & $2.40_{-0.12}^{+0.08}$& 2.93$\sim$7.89 & 60\%& 2.04\\
\hline
\end{tabular}
\par
\medskip
\begin{minipage}{0.82\linewidth}
NOTES. --- The uncertainties of $N_H$ and $kT/\alpha$ are the 90\% confidence.\\
$^a$ PL: power-law.; TB: thermal bremsstrahlung.\\
$^b$ in units of $10^{21}$ cm$^{-2}$\\
$^c$ The goodness represents the fraction of simulations with a lower C-statistic than the data. If the spectrum and the simulations were produced by the same model, there is an equal probability of obtaining any given percentage. If this fraction is very large (perhaps $\> 95$\%, or $\> 99$\%) one may conclude that it is unlikely that the data is drawn from this model.\\  
$^d$ 0.3--7 keV unabsorbed flux in units of $10^{-14}$\flux.
\end{minipage}
\end{table*}

\begin{table*}
\centering{\footnotesize
\caption{Optical Counterparts to \chandra\ X-ray Sources}
\begin{tabular}{lccccc}
\hline
\hline
Source Name  & $V_{606}$& $I_{814}$& $f_X/f_V$\tablenotemark{a} & Classification\tablenotemark{b}\\
\hline

     CX01 &    $ 23.25{\pm0.01}$ &    $ 21.59{\pm0.01}$ &   27.81 &        CV? \\
     CX02 &     $25.66{\pm0.03}$ &     $24.38{\pm0.03}$ &   23.61 &         CV? AGN? \\
     CX03a &     $18.84 {\pm0.02}$&     $17.39{\pm0.01}$ &    0.01 &         AB? \\

     CX03b &     $21.95 {\pm0.01}$&     $20.68{\pm0.01}$&    0.49 &          CV \\
     \\

     CX04a &     $18.69{\pm0.01}$&     $17.25{\pm0.01}$&    0.01 &         AB \\

     CX04b &      $22.93{\pm0.01}$ &    $ 21.33{\pm0.01}$ &     0.73 &         ? \\

     CX05a &    $ 24.14{\pm0.01}$ &     $22.30{\pm0.01}$ &    0.66 &        ? \\

     CX05b &    $ 25.56{\pm0.03}$ &    $23.41{\pm0.02}$ &    2.41 &          ? \\
     
     CX05c &  $< 18.84$ & $< 15.63$ & $< 0.01$  &  AB? \\
     \\
     CX07a &      $20.53{\pm0.01}$ &     $19.66{\pm0.01}$ &    3.40 &         CV?AGN? \\

     CX07b &     $26.16{\pm0.04}$ &     $23.11{\pm0.01}$ &    608.85 &        AGN \\

     CX08a &    $ 21.35{\pm0.05}$ &     $20.05{\pm0.08}$ &    2.457 &         ? \\

     CX08b &     $23.28{\pm0.12}$ &    $ 21.44{\pm0.12}$ &   14.40 &          CV? \\

\\
      CX28 &     $24.43{\pm0.016}$&    $ 22.00{\pm0.009}$ &    2.11 &        AGN \\

   E3-CX02 &      $22.79{\pm0.03}$ &    $22.21{\pm0.02}$ &    2.37 &         CV? AGN? \\

   E3-CX03 &     $25.06{\pm0.09}$ &     $23.09{\pm0.03}$ &    17.02 &        AGN \\

\hline
\end{tabular}
\par
\medskip
\begin{minipage}{0.97\linewidth}

$^a$ Ratio of X-ray to optical ($V_{606}$) flux, using $log
  (f_X/f_V)=log f_X + 5.67 + 0.4 V_{606}$ (Green et al. 2004); $f_X$
  is derived in the 0.3--7 keV band.\\
$^b$ CV: cataclysmic variable; AB: chromospherically active binary;
  AGN: active galactic nuclei. We consider all optical sources inside X-ray error circle as counterparts. 
\end{minipage}
}
\end{table*}

\begin{table*}
\centering{\footnotesize
\caption{Optical Counterparts Observed by \hst\ WFPC2}
\begin{tabular}{lccccc}
\hline
\hline
Source Name  & $U_{336}$& $B_{439}$& $H\alpha $ & $R_{675}$\\
\hline

     CX03a &     $19.52 {\pm0.02}$&     $19.23{\pm0.02}$ &    $17.80{\pm0.02}$ & $18.35{\pm0.02}$ \\

     CX03b &     $22.81 {\pm0.12}$&     $22.74{\pm0.13}$ &   $21.13{\pm0.11}$  &     $21.45{\pm0.10}$ \\

     CX05a &    $>24$ &     $>24$ &    $>23$ &        $>23$ \\

     CX05b &    $>25$ &    $>25$  &    $>24$ &         $>24$ \\
     CX05c &    $<17.96$ &    $<17.51$  &    $<16.24$ &         $<17.76$ \\

\hline
\end{tabular}
\par
\medskip
\begin{minipage}{0.97\linewidth}
NOTES. --- CX05a and CX05b are too faint to provide reliable magnitude.

\end{minipage}
}
\end{table*}

\begin{table*}
\centering{\footnotesize
\caption{The Probability of Positional Coincidence}
\begin{tabular}{lccc}
\hline
\hline
Source Name  & Average Number\tablenotemark{a}& Probability \tablenotemark{b} (\%)\\
\hline

CX01	&	0.375	&	31.27 	\\
CX02	&	0.375	&	31.27 	\\
CX03	&	0.5	&	9.00 	\\
CX04	&	0.375	&	5.67 	\\
CX05	&	1.25	&	15.49 	\\
CX06	&	0.625	&	-- 	\\
CX07	&	0.25	&	2.65 	\\
CX08	&	0.125	&	0.89 	\\
CX28	&	0.5	&	43.12 	\\
CX30	&	0.75	&	-- 	\\
E3-CX01	&	0.125	&	11.76 	\\
E3-CX02	&	0.125	&	11.76 	\\
E3-CX03	&	0.125	&	11.76 	\\

\hline
\end{tabular}
\par
\medskip
\begin{minipage}{0.97\linewidth}

$^a$ The error circle on the HST images of NGC6144 and E3 are shifted to eight directions  ($5''$ for NGC6144 and $10''$ for E3). Here we list the average number of sources found in the circle over 8 trials.\\
$^b$ The probability of finding the observed number (or more) sources inside the error circle, assuming Poisson distribution with a mean given by the average number. 
\end{minipage}
}
\end{table*}


\begin{references}

\reference{} Alcaino. G.\ 1980, A\&A 39, 315
\reference{} Bailyn, C. D., Grindlay, J. E., \& Garcia, M. R. \ 1990, \apj, 357, L35
\reference{} Bassa, C., et al.\ 2004, \apj, 609, 755
\reference{} Bassa, C. G., Pooley, D., Verbunt, F., Homer, L., Anderson, S. F., \& Lewin, W. H. G. \ 2008, A\&A, 488, 921B

\reference{} Clark, G.~W., Markert, T.~H., \& Li, F. K.\ 1975, \apj, 199, L93
\reference{} Cool, A.~M., Grindlay, J.~E., Cohn, H.~N., Lugger, P.~M., \& Slavin, S. D.\ 1995, \apj, 508, L75

\reference{} Cool, A. M., Haggard, D., \& Carlin, J. \ 2002, in Omega Centauri, \textsl{ A Unique Window into Astrophysics}, eds. F. van Leeuwen, J. Hughes, \& G. Piotto, 265, 277 

 

\reference{} Davies, M.~B. 1997, MNRAS, 288, 117
\reference{} Dempsey, R. C., Linsky, J. L., Fleming, T. A., \& Schmitt, J. H. M. M.\ 1993 \apjs, 86, 599


\reference{} Dolphin, A.~E.\ 2000, \pasp, 112, 1383


\reference{} Freeman, P.~E., 
Kashyap, V., Rosner, R., \& Lamb, D.~Q.\ 2002, \apjs, 138, 185 
\reference{} Giacconi, R., Murray, S., Gursky, H., Kellogg, E., Schreier, E. \& Tananbaum. H. \ 1972, \apj, 178, 281
\reference{} Giacconi, R., Rosati, P., Tozzi, P., Nonino, M., Hasinger, G., Norman, C., Bergeron, J., Borgani, S., Gilli, R., Gilmozzi, R. \& Zheng, W. \ 2001, \apj, 551, 624
\reference{} Green, P.~J., et al.\ 2004, \apjs, 150, 43


\reference{} Girardi, L., Bressan, A., Bertelli, G., \& Chiosi, C. \ 2000, A\& AS, 141, 371
\reference{} Grindlay, J.~E., Camilo, F.,
Heinke, C.~O., Edmonds, P.~D., Cohn, H., \& Lugger, P.\ 2002, \apj, 
581, 470

\reference{} Grindlay, J.~E., Heinke, C.~O., Edmonds, P.~D., Murray, S. ~S.,\& Cool, A. M.\ 2001, \apj, 563, L53


\reference{} Harris, W. 1996, AJ, 112, 1487



\reference{} Heinke, C.~O., Grindlay, J.~E., Lugger, P.~M., Cohn,
H.~N., Edmonds, P.~D., Lloyd, D.~A., \& Cool, A.~M.\ 2003, ApJ, 598, 501

\reference{} Heinke, C.~O., Grindlay, 
J.~E., Edmonds, P.~D., Cohn, H.~N., Lugger, P.~M., Camilo, F.,
Bogdanov, S., \& Freire, P.~C.\ 2005, \apj, 625, 796 
\reference{} Hills, J. G.\ 1975, \aj, 80, 809
\reference{} Hills, J. G.\ 1976, \mnras, 175, 1P


\reference{} Hertz, P., \&  Grindlay, J.~E.\ 1983, \apj, 275, 105
\reference{} Kalberla, P.M.W., Burton, W.B., Hartmann, Dap, Arnal, E.M., Bajaja, E., Morras, R., \& Poppel, W.G.L. \ 2005, A \& A, 440, 775

\reference{} Kong, A. K. H., Bassa, C., Pooley. D., Lewin W. H. G., Homer L., Verbunt F., Anderson S. F., \& Margon B. \ 2006, \apj, 647, 1065

\reference{} Lu, T. N., Kong, A. K. H., Bassa, C., Verbunt F., Lewin W. H. G., Anderson S. F., Pooley. D.,  Homer L.,  \ 2009, \apj, 705, 175

\reference{} Nousek, J. A., \& Shue, D. R. \ 1989, \apj,  342, 1207 
\reference{} Patterson, J. \ 1984, \apjs, 54, 443
\reference{} Pooley, D., et al.\ 2002, \apj, 569, 405
\reference{} Pooley, D., et al.\ 2003, \apjl, 591, L131
\reference{} Pooley, D., \& Hut, P. \ 2006, \apj, 646, L143
\reference{} Predehl, P., \& 
Schmitt, J.~H.~M.~M.\ 1995, \aap, 293, 889

\reference{} Pryor, C., et al.\ 1991, ASPC, 13, 439p

\reference{} Sarajedini, A., et al.\ 2007,  \aj , 133, 1658 
\reference{} Servillat, M., et al.\ 2008, A \& A, 490, 641
\reference{} Verbunt, F. \ 2002, in $\omega$\,Cen, a unique window
into astrophysics, eds. F. van Leeuwen, Hughes, J. \& G. Piotto, ASP
Conf.\ Ser.\ 265, 289

\reference{} Verbunt, F.\ 2003, ASP 
Conf.~Ser.~296: New Horizons in Globular Cluster Astronomy, 296, 245

\reference{} Verbunt, F. \& Lewin, W. H. G. \ 2006, Globular cluster X-ray sources (Compact stellar X-ray sources, ed. L. Walter, \& M. van der Klis, Cambridge Astrophysics Series, No. 39 (Cambridge, UK: Cambridge University Press), 341 

\reference{} Verbunt, F., Pooley, D., \& Bassa, C. 2007, in IAU Symposium, Vol. 246, IAU Symposium, 301-310
\reference{} van den Bergh, S., Demers. S., \& Kunkel. W. E. \ 1980, \apj, 239, 122 
\reference{} Wijnanads, R., Heinke, C., Pooley, D., et al. \ 2005, \apj, 618, 883



\end{references}
\end{document}